\documentstyle[axodraw]{aipproc}

\begin{document}
\title{Heavy Flavor Contributions to QCD Sum Rules and the Running Coupling
Constant\thanks{Talk presented at the International Workshop on Hadron
Physics `Effective Theories of Low Energy QCD', Coimbra, Portugal, 10-15 
September, 1999.}}

\author{W.L.van Neerven\thanks{Work supported by the 
EC network `QCD and Particle Structure'  under contract No.~FMRX--CT98--0194.}}
\address{Instituut-Lorentz, Universiteit Leiden, P.O. Box 9506, 2300 RA Leiden,
The Netherlands.}

\maketitle

\begin{abstract}
We have calculated first and second order corrections to several
sum rules measured in deep inelastic lepton-hadron scattering. These
corrections, which are due to heavy flavors only, are compared with
the existing perturbation series which is computed for massless quarks
up to third order in the strong coupling constant $\alpha_s$. A study
of the perturbation series reveals that the large logarithms of the type
$\ln^i Q^2/m^2$ dominate the perturbation series at much larger values
than those given by the usual matching conditions imposed on the
$\alpha_s(\mu)$. Therefore these matching
conditions cannot be used to extrapolate the running coupling constant
from small $\mu$ to very large scales like $\mu=M_Z$. An alternative 
description of the running coupling constant in the MOM-scheme is proposed.
\end{abstract}

\section*{Introduction}

Structure functions, measured in deep inelastic lepton-hadron scattering
(see\\ Fig. 1)
\begin{eqnarray}
l_1(k_1) + H(p) \rightarrow l_2(k_2) + `X'\,,
\end{eqnarray}
where $`X'$ denotes any inclusive final hadronic state, provide us with an 
excellent test of perturbative QCD. The in and outgoing
leptons are represented by $l_1$ and $l_2$ respectively and the hadron is
denoted by $H$.
On the Born level the reaction proceeds via the exchange of one of
the vector bosons $V$ of the standard model which are given by
$\gamma$, $Z$ and $W^{\pm}$. The virtuality of the vector boson $V$ and
the C.M. energy are given by $q^2=-Q^2 < 0$ and $S=(p+k_1)^2$ respectively. 
Further the scaling variables are defined by $x=Q^2/2p.q$ and $y=p.q/p.k_1$.
\begin{figure}
\begin{center}
  \begin{picture}(185,135)(0,0)
    \ArrowLine(0,20)(70,30)
    \Line(90,30)(148,35)
    \Line(80,25)(160,25)
    \Line(90,20)(148,15)
    \Line(130,50)(160,25)
    \Line(130,0)(160,25)
    \GCirc(80,30){20}{0.3}
    \Photon(50,100)(72,48){3}{7}
    \ArrowLine(0,100)(50,100)
    \ArrowLine(50,100)(100,120)
    \Text(0,115)[t]{$l_1$}
    \Text(100,135)[t]{$l_2$}
    \Text(0,15)[t]{$H$}
    \Text(50,75)[t]{$V$}
    \Text(175,30)[t]{$`X'$}
    \Text(30,20)[t]{$p$}
    \Text(25,115)[t]{$k_1$}
    \Text(75,125)[t]{$k_2$}
    \Text(75,75)[t]{$\downarrow q$}
  \end{picture}
  \caption[]{Kinematics of deep inelastic lepton-hadron scattering.}
  \label{fig1}
\end{center}
\end{figure}
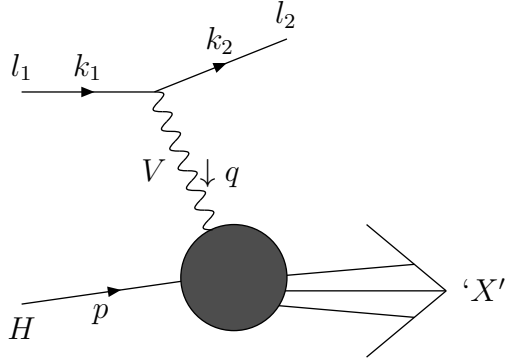
The structure functions show up in the polarized and unpolarized cross 
sections. Starting with the spin-averaged cross section for $V=\gamma$ we
obtain
\begin{eqnarray}
\frac{d^2\sigma}{d\,x\,d\,y}=\frac{4\pi\alpha^2}{(Q^2)^2} S \left [
(1-y) F_2(x,Q^2)+ x\,y^2  F_1(x,Q^2) \right ] \,.
\end{eqnarray}
For $V=W^{\pm}$ and $Q^2 \ll M_W^2$ the unpolarized cross section for the 
charged current process becomes
\begin{eqnarray}
\frac{d^2\sigma^{\pm}}{d\,x\,d\,y}=\frac{G_F^2}{2\pi} S \left [
(1-y) F_2(x,Q^2)+ x\,y^2  F_1(x,Q^2) \mp x\,y\,(1-\frac{y}{2}) F_3(x,Q^2)
\right ] \,.
\end{eqnarray}
Finally we are also interested in the spin structure functions which are
measured in polarized scattering. In the case of $V=\gamma$ the cross section
becomes
\begin{eqnarray}
\frac{d^2\sigma^{{\rm H}(\rightarrow)}(\rightarrow)}{d\,x\,d\,y}
-\frac{d^2\sigma^{{\rm H}(\leftarrow)}(\rightarrow)}{d\,x\,d\,y}
=\frac{8\pi\alpha^2}{Q^2} \left [ (2-y) g_1(x,Q^2) \right ]\,,
\end{eqnarray}
from which one can extract the longitudinal spin structure function $g_1$.
The transverse spin structure function $g_2$ is obtained from the
cross section
\begin{eqnarray}
\frac{d^3\sigma^{{\rm H}(\uparrow)}(\rightarrow)}{d\,x\,d\,y\,d\,\phi}
&-&\frac{d^3\sigma^{{\rm H}(\downarrow)}(\rightarrow)}{d\,x\,d\,y\,d\,\phi}
\nonumber\\[2ex]
&=&\cos \phi\frac{4\alpha^2}{Q^2} \left (\frac{4M^2x(1-y)}{yS} \right )^{1/2}
\left [ y g_1(x,Q^2)+ 2 g_2(x,Q^2) \right ] \,,
\end{eqnarray}
where $\phi$ is the angle between the spin of the hadron and the plane spanned
by the lepton momenta $\vec k_i$. Further we have indicated between the brackets
the spin of the incoming lepton $l_1$ and we have neglected in the expressions
above all power corrections of the type $M^2/S$ where $M$ denotes the mass of
the hadron $H$ (see Fig. \ref{fig1}).

One of the most important predictions of QCD is the $Q^2$-evolution of
the structure functions. However the $x$-dependence cannot be predicted yet.
The latter follows from non-perturbative QCD which is at such a premature stage
that the $x$-dependence cannot be determined. This problem can be avoided
when one integrates over the latter variable which leads to the sum rules 
discussed below. Notice that this integration requires a full knowledge
about the range $0<x<1$. Since there are no data available for 
$x<10^{-4}$ and $x>0.8$ one has to make extrapolations into this region. 
This will introduce an error which is very hard to estimate.
In the past various techniques have been used to compute the sum rules. The 
most known among them are
\begin{itemize}
\item[1.] Infinite momentum frame techniques used in current algebra.
\item[2.] Dispersion relations which are derived from the Regge behavior
of the structure functions $F(x,Q^2)$ at small $x$.
\item[3.] Operator product expansion (OPE) techniques which in leading twist
are equivalent to the parton model.
\end{itemize}
Using the last technique the derivation of the sum rules proceeds as follows.
First let us define the hadronic structure tensor which appears in the 
derivation of the cross sections presented above. It is given by
\begin{eqnarray}
 W^{\mu\nu}(p,q,s)&=&\frac{1}{4\pi} \int d^4z \, e^{iq.z} \langle p,s \mid
[J^{\mu}(z),J^{\nu}(0)] \mid p,s \rangle
\nonumber\\[2ex]
&=&W_S^{\mu\nu}(p,q) +iW_A^{\mu\nu}(p,q,s)\,.
\end{eqnarray}
This tensor can be split up into a symmetric and a antisymmetric part i.e.
\begin{eqnarray}
W_S^{\mu\nu}(p,q)&=&\left (-g^{\mu\nu}+\frac{q^{\mu}q^{\nu}}{q^2} \right )
F_1(x,Q^2) 
\nonumber\\[1ex]
&& +\left (p^{\mu}-\frac{p\,q}{q^2}q^{\mu}\right )\left (p^{\mu}
-\frac{p\,q}{q^2}q^{\mu} \right ) \frac{F_1(x,Q^2)}{p\,q} \,,\\[2ex]
W_A^{\mu\nu}(p,q,s)&=&-\frac{M}{p\,q}\epsilon^{\mu\nu\alpha\beta}q_{\alpha}
\left [s_{\beta}g_1(x,Q^2)+(s_{\beta}-\frac{s\,q}{p\,q}p_{\beta})g_2(x,Q^2) 
\right ] \,.
\end{eqnarray}
In the Bjorken limit ($Q^2 \rightarrow \infty$ and $x$ is fixed) the
integrand in Eq. (6) is dominated by the lightcone $z^2=0$. This allows us
to make a lightcone expansion of the current commutator
\begin{eqnarray}
 [J(z),J(0)]
 {\raisebox{-2 mm}{$\,\stackrel{=}{{\scriptstyle z^2 \sim 0 }}\, $} }
\sum_{\tau} \sum_{N} C^{N,\tau}(z^2 \mu^2) O^{N,\tau}(\mu^2,0)\,,
\end{eqnarray}
where for convenience we have dropped the Lorentz indices of the currents.
In the expression above $\tau$ and $N$ denote the twist and spin of the
operator $O^{N,\tau}$ respectively. The latter and the singular coefficient
function ${\cal C}^{N,\tau}$ are understood to be renormalized where $\mu$
represents the renormalization scale. After insertion of expression (11) into
Eq. (6) one obtains the $N$th moment of the structure function which equals
\begin{eqnarray}
 \int_0^1dx\,x^{N-1}\,F(x,Q^2)= \sum_{\tau} \left (\frac{M^2}{Q^2}
\right )^{\frac{\tau}{2}-1} A^{(N),\tau}(\mu^2)\,{\cal C}^{(N),\tau}
\left(\frac{Q^2}{\mu^2}\right) \,.
\end{eqnarray}
In momentum space the operator matrix element (OME) and the coefficient 
function are defined by
\begin{eqnarray}
A^{(N),\tau}(\mu^2)=\langle p \mid O^{N,\tau}(\mu^2,0) \mid p \rangle \,,
\end{eqnarray}
and
\begin{eqnarray}
{\cal C}^{(N),\tau}\left(\frac{Q^2}{\mu^2}\right)=\int d^4 z \,e^{iq.z}
C^{N,\tau}(z^2 \mu^2) \,,
\end{eqnarray}
respectively. Limiting ourselves to twist ${\it two}$ contributions only
the non-singlet part (w.r.t flavor) of the structure functions is determined
by the quark operator
\begin{eqnarray}
O^{\rm N}_k \equiv O_k^{\mu_1\mu_2 \cdots \mu_N}(x)= 
\bar \psi(x) \gamma^{\mu_1} D^{\mu_2} \cdots  D^{\mu_2} \lambda_k \psi (x)\,,
\end{eqnarray}
where $D^{\mu}$ denotes the covariant derivative and $\lambda_k$ are the 
generators of the flavor group $SU_F(n_f)$. In the case of $N=1$ the following
operators are conserved
\begin{eqnarray}
O_k^{\mu}(x)=\bar \psi(x)\gamma^{\mu}\lambda_k\psi(x) \quad \mbox{and}\,, \quad
O_k^{\mu,5}(x)=\bar \psi(x)\gamma^{\mu}\gamma^5\lambda_k\psi(x) \,,
\end{eqnarray}
which means that they do not have to renormalized. Hence the corresponding OME's
and coefficient functions are independent of the scale $\mu$ i.e.
\begin{eqnarray}
 \int_0^1dxF(x,Q^2)= A_k^{(1)} {\cal C}_k^{(1)} (Q^2) \,, \quad
\mbox{with} \quad \langle p \mid O_k^{\mu}(0) \mid p \rangle =  A_k^{(1)}p^{\mu}
\,,
\end{eqnarray}
where $A_k^{(1)}$ can be determined from $SU_F(n_f)$ and a low energy 
theorem.
In general the coefficient function ${\cal C}^{(1)}$ can be expanded in a
power series in $\alpha_s(\mu^2)$. However in two cases this function
becomes trivial. The first example is the Adler sum rule \cite{ad}
\begin{eqnarray}
&& \int_0^1\, \frac{dx}{x} \,\left [F_2^{\bar \nu p}(x,Q^2)
- F_2^{\nu p}(x,Q^2) \right ] = K(n_f) \,,
\nonumber\\[2ex]
&& K(3) = 2 + 2 \sin^2 \theta_c \quad (SU_F(3)) \,, \qquad K(4) =
2\quad (SU_F(4)) \,.
\end{eqnarray}
Here $\theta_c$ denotes the Cabibbo angle and for the constant $K(n_f)$
we have quoted the values given by the flavor group $SU_F(n_f)$ for 
$n_f=3,4$. The second example is the Burkhardt-Cottingham sum rule \cite{buha}
given by
\begin{eqnarray}
\int_0^1 dx\, g_2(x,Q^2)=0\,.
\end{eqnarray}
The above sum rules above have the following properties
\begin{itemize}
\item[1.] The values on the righthand side are independent of the method used
for the derivation of the sum rules.
\item[2.] No power corrections of the type $(\Lambda^2/Q^2)^i$ (higher twist).
\item[3.] No mass corrections (e.g. due to heavy flavors).
\item[4.] No QCD corrections.
\end{itemize}
The coefficient function has a nontrivial form in the following cases
provided the results follow from OPE. The first one is the polarized Bjorken
sum rule \cite{bjork1}
\begin{eqnarray}
\Delta g_1(Q^2)\equiv \int_0^1\, dx\, \left [g_1^{ep}(x,Q^2) -
g_1^{en}(x,Q^2)
\right ] = \frac{1}{6} \left| \frac{G_A}{G_V} \right| {\cal C}^{g_1}(n_f,Q^2)\,.
\end{eqnarray}
The second one is represented by the unpolarized Bjorken sum rule \cite{bjork1}
\begin{eqnarray}
 \Delta F_1(Q^2)& \equiv& \int_0^1\, dx\, \left [F_1^{\bar \nu p}(x,Q^2)
- F_1^{\nu p}(x,Q^2) \right ] = K(n_f) \, {\cal C}^{F_1}(n_f,Q^2)\,,
\nonumber\\[2ex]
 K(3) &=& 1 + \sin^2 \theta_c \quad ( SU_F(3)) \,,
\qquad K(4) = 1 \quad (SU_F(4)) \,,
\end{eqnarray}
The third one is the Gross-Llewellyn Smith sum rule \cite{grls}
\begin{eqnarray}
\Delta F_3(Q^2)&\equiv& \int_0^1\, dx\, \left [F_3^{\bar \nu p}(x,Q^2)
+ F_3^{\nu p}(x,Q^2) \right ] = K(n_f) \, {\cal C}^{F_3}(n_f,Q^2)\,,
\nonumber\\[2ex]
 K(3) &=& 6 - 2 \sin^2 \theta_c \quad ( SU_F(3)) \,, \qquad K(4) = 6 \quad
(SU_F(4)) \,.
\end{eqnarray}
The coefficient function can be expanded in $\alpha_s$ as
\begin{eqnarray}
{\cal C}^{r}(n_f,Q^2)=\sum_{i=0}^{\infty} \left (\frac{\alpha_s(n_f,Q^2)}{\pi}
\right )^i c_i^{r}(n_f)\,, \quad r=g_1,F_1,F_3 \,.
\end{eqnarray}
The sum rules in Eqs. (18)-(20) have the following properties
\begin{itemize}
\item[1.] For $m_q=0$ the $c_i^r$ are known up to order $\alpha_s^3$ for 
$i\le 3$ (see \cite{latk}, \cite{lave}).
\item[2.] Higher twist corrections are estimated in \cite{babr}
\item[3.] Heavy flavor corrections are computed up to order $\alpha_s^2$
in \cite{blne}
\end{itemize}
In leading twist the coefficient functions can be computed in the QCD
improved parton model. In this model the structure function can be written as
\begin{eqnarray}
F(x,Q^2)&=&\sum_a e_a^2 \int_0^1 dz_1 \,\int_0^1 dz_2\, \delta(x-z_1z_2) \hat
f_a^{\rm H}(z_1) \, \hat {\cal F}_a (z_2,\frac{Q^2}{m_q^2})
\nonumber\\[2ex]
&\equiv& \sum_a e_a^2  \hat f_a^{\rm H} \otimes {\cal F}_a (\frac{Q^2}{m_q^2})
\,,
\end{eqnarray}
where $\hat f_a^{\rm H}$ denotes the bare parton density and the parton
structure function ${\cal F}_a$ represents the QCD corrections including the 
Born approximation. Notice that ${\cal F}_a$ contains mass singularities.
They are regulated by giving the light quark a mass which is sufficient in the 
non-singlet case. Subsequently these singularities are removed via mass
factorization which reads as follows
\begin{eqnarray}
{\cal F}_a (\frac{Q^2}{m_q^2})=A_{aa}(\frac{\mu^2}{m_q^2}) \otimes 
{\cal C}_a (\frac{Q^2}{\mu^2})\,.
\end{eqnarray}
Hence the hadronic structure function can be written as
\begin{eqnarray}
F(Q^2)=\sum_a e_a^2 f_a^{\rm H}(\mu^2) \otimes {\cal C}_a (\frac{Q^2}{\mu^2})
\,, \quad \mbox{with} \quad f_a^{\rm H}(\mu^2)=A_{aa}(\frac{\mu^2}{m_q^2}) 
\otimes \hat f_a^{\rm H}\,,
\end{eqnarray}
where $f_a^{\rm H}$ denotes the renormalized parton density which like
the coefficient function depends on the factorization scale $\mu$. Performing
the Mellin transformation one obtains
\begin{eqnarray}
\int_0^1 dx\,x^{N-1}\,F(x,Q^2)=A^{(N)}(\mu^2)\,{\cal C}^{(N)}(\frac{Q^2}{\mu^2})
\,,
\end{eqnarray}
so that one can make the following identifications
\begin{eqnarray}
A^{(N)}(\mu^2)=\int_0^1 dz\,z^{N-1}\,\sum_a e_a^2 f_a^{\rm H}(z,\mu^2) \,,
\quad {\cal C}^{(N)} (\frac{Q^2}{\mu^2})=\int_0^1 dy\,y^{N-1}\, 
{\cal C}(y,\frac{Q^2}{\mu^2})\,,
\end{eqnarray}
which yields the coefficient function ${\cal C}^{(1)}$
\section*{Heavy flavor thresholds in the strong coupling constant}
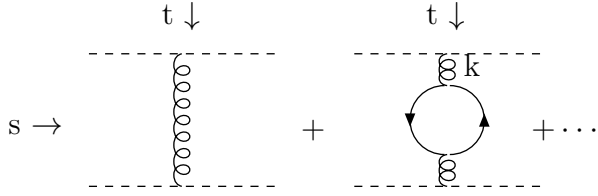
\begin{figure}
\begin{center}
  \begin{picture}(200,100)(0,0)
    \DashLine(20,10)(90,10){3}
    \DashLine(20,60)(90,60){3}
    \Gluon(55,10)(55,60){3}{7}
    \DashLine(120,10)(190,10){3}
    \DashLine(120,60)(190,60){3}
    \Gluon(155,10)(155,22){3}{2}
    \Gluon(155,48)(155,60){3}{2}
    \ArrowArc(156,35)(13,270,90)
    \ArrowArc(154,35)(13,90,270)
    \Text(105,35)[t]{$+$}
    \Text(0,35)[t]{s $\rightarrow$}
    \Text(55,80)[t]{t $\downarrow$}
    \Text(155,80)[t]{t $\downarrow$}
    \Text(165,60)[t]{k}
    \Text(200,35)[t]{$+ \cdots$}
  \end{picture}
  \caption[]{heavy flavor loop contribution to quark-quark scattering.}
  \label{fig2}
\end{center}
\end{figure}
Coupling constant renormalization in the case of light flavors (massless
quarks) conventionally proceeds in the ${\overline {\rm MS}}$-scheme. Using
n-dimensional regularization the coupling constant gets a dimension
i.e. $g_s^u\,M^{4-n}$ where $g_s^u$ is the bare (unrenormalized) coupling
constant. The renormalization of the perturbation series is then carried
out by the substitution
\begin{eqnarray}
g_s^u \rightarrow g_s(\mu^2) \left [1 + \frac{\alpha_s(\mu^2)}{8\pi} 
\beta_0(n_f) \left \{ \frac{2}{\varepsilon} +\gamma_E - \ln 4\pi + 
\ln \frac{\mu^2}{M^2} \right \} + \cdots \right ] \,,
\end{eqnarray}
with $\alpha_s=g_s^2/4 \pi$ and $\beta_0(n_f)=11-2/3\,n_f$. The renormalized 
coupling constant $\alpha_s$ satisfies the differential equation
\begin{eqnarray}
\mu^2 \frac {d\,\alpha_s(n_f,\mu^2)}{d\,\mu^2}=\beta(\alpha_s(n_f,\mu^2)) 
\,, \quad \rightarrow \quad
\alpha_s(n_f,\mu^2)= \frac{4\pi}{\beta_0(n_f)\ln \frac{\mu^2}{\Lambda^2}} 
+ \cdots \,.
\end{eqnarray}
The solution obtained from the equation above is quite appropriate since the 
scale $\mu$ is much larger than the light quark mass. However in the case of 
heavy quarks we encounter a problem which will be illustrated by the following
example. Let us consider (light) quark-quark scattering as shown in Fig. 
\ref{fig2}. Including the heavy quark loop the cross section behaves like
\begin{eqnarray}
\sigma \sim \alpha_s^2(n_f,\mu^2) \frac{1}{t} + 2\alpha_s^3(n_f,\mu^2)
\frac{1}{t} \Pi_1 (\frac{t}{m^2}) + \cdots \,,
\end{eqnarray}
where $\Pi_1(x)$ denotes the one-loop vacuum polarization function which 
behaves asymptotically as
\begin{eqnarray}
 \Pi_1 \left (\frac{t}{m^2} \right )  
{\raisebox{-2 mm}{$\,\stackrel{\sim}{{\scriptstyle -t \gg m^2 }}\, $} }
- \beta_{0,Q} \ln \left (\frac{-t}{m^2} \right )\,,
\qquad \beta_{0,Q}=-\frac{2}{3}\,.
\end{eqnarray}
This behavior leads to large corrections in the perturbation series which
have to be resummed as follows. First we have to make the substitution
\begin{eqnarray}
 \alpha_s(n_f,\mu^2) \rightarrow  \alpha_s(n_f+1,\mu^2) [1 +
 \alpha_s(n_f+1,\mu^2) \beta_{0,Q} \ln \frac{\mu^2}{m^2} \cdots ]\,,
\end{eqnarray}
in Eq. (29) so that for $-t \gg m^2$ the cross section behaves like
\begin{eqnarray}
\sigma \sim \alpha_s^2(n_f+1,\mu^2) \frac{1}{t} + 2\alpha_s^3(n_f+1,\mu^2)
\frac{1}{t} \beta_{0,Q} \ln \frac{\mu^2}{-t}\,.
\end{eqnarray}
Next $\alpha_s(n_f+1,\mu^2)$ satisfies the massless renormalization group
equation in (28) but now for $n_f+1$ light flavors with the boundary condition
\cite{ma}
\begin{eqnarray}
\alpha_s(n_f,\Lambda_{n_f},\mu^2)=\alpha_s(n_f+1,\Lambda_{n_f+1},\mu^2)\,,
\end{eqnarray}
where the matching value is given by $\mu=m$. In this way one obtains
a better behavior of the perturbation series than shown by Eq. (29) in 
particular for $\mu^2=-t$. However the representation of the cross section 
in (32) at $n_f+1$ light flavors is only correct if $-t\gg m^2$. This follows
from the vacuum polarization function $\Pi_1(-t/m^2)$ which gets close to the 
logarithm in Eq. (30) for $-t>21~m^2$ within $10\%$. For $-t\approx m^2$
one does not reach the asymptotic form of $\Pi_1(-t/m^2)$ so that
in this case it is much better to use the representation for the cross section
at $n_f$ flavors in Eq. (29). It also shows that the matching scale $\mu$
in Eq. (31) has to chosen at a much larger value e.g. $\mu^2=21~m^2$
instead of $\mu^2=m^2$.

\section*{Heavy Flavor contributions to the polarized Bjorken sum rule}

\begin{table}[b!]
\caption{QCD corrections to the polarized Bjorken sum rule.}
\label{table1}
\begin{tabular}{ldddd}
$Q^2$ $({\rm GeV/c})^2$ & light flavors&
   \multicolumn{1}{c}{charm contribution\tablenote{Exact formula Eq. (35).}} &
  \multicolumn{1}{c}{charm contribution\tablenote{Asymptotic formula
according to Eq. (37).}} &  \multicolumn{1}{c}{charm contribution\tablenote
{Estimated order $\alpha_s^2$ correction according to \cite{kast}.}}\\
\tableline
3  & 0.795 & -0.688.$10^{-3}$ & 0.451.$10^{-2}$ & -0.265.$10^{-1}$\\
10  & 0.883 & -0.788.$10^{-3}$ &  0.569.$10^{-3}$ &  -0.586.$10^{-2}$ \\
100  & 0.931 & -0.107.$10^{-2}$ &  -0.0912.$10^{-2}$ & - 0.121.$10^{-2}$\\
\end{tabular}
\end{table}
The sum rule in Eq. (18) receives contributions from the following subprocesses
(see Fig. 2 in \cite{blne})
\begin{eqnarray}
&& q + \gamma^* \rightarrow q + Q + \bar Q \,,
\nonumber\\[1ex]
&& q + \gamma^* \rightarrow q + g \quad \mbox{one-loop gluon self energy 
correction} \,,
\nonumber\\[1ex]
&& q + \gamma^* \rightarrow q \quad \mbox{two-loop vertex correction with 
one-loop gluon self energy}\,.
\end{eqnarray}
The correction to the sum rule up to order $\alpha_s^2$ can be expressed
as
\begin{eqnarray}
{\cal C}^{g_1}(n_f,Q^2)&=&1 + \frac{\alpha_s(n_f,\mu^2)}{\pi} a_1 +
\left (\frac{\alpha_s(n_f,\mu^2)}{\pi}\right )^2 \left [ -a_1\beta_0(n_f)\ln
\frac{Q^2}{\mu^2} + a2 + n_f b_2 \right.
\nonumber\\[2ex]
&& \left. + H_Q(\frac{Q^2}{m^2}) \right ]\,,
\end{eqnarray}
where $H_Q$ denotes the heavy quark contribution. The two regions
of interest are
\begin{eqnarray}
&& Q^2 \ll m^2 \rightarrow H_Q\left (\frac{Q^2}{m^2} \right ) \sim 
\frac{Q^2}{m^2} \ln \frac{Q^2}{m^2}\,,\\   
&& Q^2 \gg m^2 \rightarrow H_Q \left (\frac{Q^2}{m^2}\right ) = -a_1 
\beta_{0,Q} \ln \frac{Q^2}{m^2} +b_2 \,.
\end{eqnarray}
The first limit shows that heavy quarks of mass $m$ decouple from the 
perturbation series when their mass is much larger than the virtuality
of the photon. In the region $Q^2 \gg m^2$ we observe the same logarithmic
behavior as for $\Pi_1$ in Eq. (30). However here ${\cal C}^{g_1}(n_f,Q^2)$
becomes asymptotic for $Q^2 >40~m^2$ within $15\%$. Hence the value 
of $Q^2$ is about twice as large as the one
found for $-t$ in quark-quark scattering. This we have checked for
charm production. Here we choose $n_f=3$ and $m=m_c=1.5~{\rm GeV/c^2}$ in 
Eq. (35) and the results are shown in table 1 (see \cite{blne}).
Further we observe that the heavy flavor corrections to the sum rule (21)
are very small. They are even smaller than the estimated order $\alpha_s^4$
corrections in \cite{kast} in spite of the fact that the heavy
flavor contribution $H_Q$ starts in order $\alpha_s^2$. The reason that the
logarithmic behavior in process (36) emerges at a much larger scale than
the reaction in Fig. 2 can be explained as follows. In Fig. 2 the gluon
momentum $k$ entering the self energy is an external variable since $t=k^2$. 
However in the case of process (36) one has to integrate over $k$
so that the latter is not an  external variable anymore. Therefore the role 
of $t$ is taken over by $Q^2$. At very large $Q^2$ the perturbation series
in Eq. (35) can be improved by following the same procedure as
outlined below Eq. (30). The result is that instead of Eq. (35) one obtains
the improved perturbation series
\begin{eqnarray}
{\cal C}^{g_1}(n_f+1,Q^2)&=&1 + \frac{\alpha_s(n_f+1,\mu^2)}{\pi} a_1 +
\left (\frac{\alpha_s(n_f+1,\mu^2)}{\pi}\right )^2 \left [ -a_1\beta_0(n_f+1)
\ln \frac{Q^2}{\mu^2} \right.
\nonumber\\[2ex]
&& \left. + a2 + (n_f+1) b_2 \right ] \,,
\end{eqnarray}
which behaves much better for the choice $\mu^2=Q^2$ provided $Q^2 \gg m^2$. 
However the expression above is not a correct representation when 
$Q^2 \approx m^2$ 
as we have already seen below Eq. (33). Hence $\mu^2=m^2$ is not
a good matching scale where the running coupling constant jumps from $n_f$
to $n_f+1$ flavors. Therefore this scale has to be chosen at a much larger 
value than what is usually done in the literature.
This also means that extrapolations from a measurement of $\alpha_s(n_f,\mu^2)$
at small $\mu$ to large values like $\mu=M_Z$ have to be distrusted.
A scheme (hereafter called MOM) which incorporates the effect of the heavy 
flavor thresholds in the running coupling constant in a much better way is 
proposed in \cite{sh}. Here one simply resums the vacuum polarization function
as follows
\begin{eqnarray}
\alpha_s^{\rm MOM}(\mu^2)=\frac{\alpha_s(3,\mu^2)}{1+
\frac{\alpha_s(3,\mu^2)}{\pi} U_1 +
\frac{\alpha_s(3,\mu^2)}{\pi} \Big (U_2/U_1 \Big ) \ln \left ( 1 +
\frac{\alpha_s(3,\mu^2)}{\pi} U_1 \right ) } \,,
\end{eqnarray}
with
\begin{eqnarray}
 U_i= \sum_{f=c,b,t} \left [\Pi_i \left (\frac{\mu^2}{m_f^2}\right )-
\Pi_i \left (\frac{\mu_0^2}{m_f^2}\right ) \right ] \,, \qquad i=1,2 \cdots \,,
\end{eqnarray}
where the coupling constant $\alpha_s(3,\mu^2)$ is represented in the 
${\overline {\rm MS}}$-scheme.  Expression (39) agrees rather well with the 
numerical 
solution of the renormalization group equation in Eq. (28) where the
$\beta$-function depends on $m$. This in particular holds when 
the two-loop self energy contributions $\Pi_2$  are included.
From the results discussed above we have seen that the scales where both 
$\Pi_1$ and $H_Q$ behave logarithmically are relatively much closer to each 
other than to $\mu=m$. Hence it is better to put 
$n_f=3$ in Eq. (35) and to make the replacement
\begin{eqnarray}
\alpha_s(3,\mu^2)=\alpha_s^{\rm MOM}(\mu^2) \left [1 +
\frac{\alpha_s^{\rm MOM}(\mu^2)}{\pi} U_1 + \cdots \right ]\,,
\end{eqnarray}
than to follow the procedure outlined below Eq. (30).
In this way one gets a continuous transition through the flavor thresholds, 
also with respect to derivatives, of the perturbation series of 
${\cal C}^{g_1}$ in Eq. (35). 
Similar results were also found for the sum rules in Eqs. (19)
and (20) (see \cite{blne}). Finally we conclude
\begin{itemize}
\item[1.] Heavy flavor corrections to the sum rules are very small and they
do not exceed the estimated order $\alpha_s^4$ contributions due to light
quarks.
\item[2.] The matching scale in the literature (here $\mu=m$) where the 
running coupling constant jumps from $n_f$ to $n_f+1$ flavors is chosen to be 
too small. 
\item[3.]
The MOM scheme as chosen in Eq. (39) gives a better representation for the 
perturbation series than the usual ${\overline {\rm MS}}$-scheme with the 
matching condition in Eq. (33).
\end{itemize}

\end{document}